# Focused time-lapse inversion of radio and audio magnetotelluric data

Marina Rosas Carbajal[1], Niklas Linde[1] and Thomas Kalscheuer[2]

[1]Institute of Geophysics, University of Lausanne, Lausanne, Switzerland.

[2]Institute of Geophysics, ETH Zurich, Zurich, Switzerland.

**Corresponding author:**

Rosas Carbajal, Marina

Marina.Rosas@Unil.ch

Bâtiment Amphipôle

UNIL Sorge

CH-1015 Lausanne

Switzerland

Tél: +41 21 692 44 13

Fax: +41 21 692 44 05

**Abstract**

Geoelectrical techniques are widely used to monitor groundwater processes, while surprisingly few studies have considered audio (AMT) and radio (RMT) magnetotellurics for such purposes. In this numerical investigation, we analyze to what extent inversion results based on AMT and RMT monitoring data can be improved by (1) time-lapse difference inversion; (2) incorporation of statistical information about the expected model update (i.e., the model regularization is based on a geostatistical model); (3) using alternative model norms to quantify temporal changes (i.e., approximations of $l_1$ and Cauchy norms using iteratively reweighted least-squares), (4) constraining model updates to predefined ranges (i.e., using Lagrange Multipliers to only allow either increases or decreases of electrical resistivity with respect to background conditions). To do so, we consider a simple illustrative model and a more realistic test case related to seawater intrusion. The results are encouraging and show significant improvements when using time-lapse difference inversion with non $l_2$ model norms. Artifacts that may arise when imposing compactness of regions with temporal changes can be suppressed through inequality constraints to yield models without oscillations outside the true region of temporal changes. Based on these results, we recommend approximate $l_1$-norm solutions as they can resolve both sharp and smooth interfaces within the same model.

## 1. Introduction

Reliable monitoring of fluid redistribution and mass transfer in the subsurface are key elements to maximize oil, gas, and geothermal production, to evaluate the performance of $CO_2$ sequestration, or to manage environmental risk, such as saltwater infiltration in coastal areas. Time-lapse inversions of geophysical data enable subsurface monitoring and have been explored widely for diverse applications using a range of geophysical techniques. Time-lapse inversions resolve temporal changes better than differencing models from separate inversions because of enhanced cancellation of errors that are constant over time and because the model regularizations can be defined with respect to temporal changes. For example, LaBrecque and Yang (2001) showed that time-lapse difference inversion of 3D electrical resistance tomography (ERT) data yield models with increased error cancellation, faster convergence and higher resolution with fewer artifacts compared to differencing of separately inverted models. Ajo-Franklin et al. (2007) inverted temporal differences in crosshole seismic traveltimes to better resolve subsurface variations related to $CO_2$ sequestration. Doetsch et al. (2010) jointly inverted time-lapse crosshole electrical resistance and ground penetrating radar traveltime data to obtain improved images of moisture content plumes.

The last twenty years have seen tremendous advances in the use of geophysics for inferring temporal changes in groundwater systems (e.g., Rubin and Hubbard, 2005), but surprisingly few published studies consider inductive electromagnetic techniques (e.g., Falgàs et al., 2009; Minsley et al., 2011). This is even more puzzling given the very long tradition of inductive methods in groundwater resources evaluations (e.g., Fitterman and Stewart, 1986; Tezkan, 1999; d'Ozouville et al., 2008). One possible reason relates to the success and flexibility of ERT for this type of applications (e.g., Kemna et al., 2002). Nevertheless, inductive techniques, such as radio magnetotellurics (RMT), have some distinct advantages compared with ERT: they are more sensitive to conductors that often represent the monitoring target; they work well in regions of high contact resistance (e.g., Beylich et al., 2003); they are better suited for investigating anisotropy (Linde and Pedersen, 2004a); and they might provide models with superior resolution (for conductive structures) compared with ERT (Kalscheuer et al., 2010). The same properties hold for audio magnetotelluric (AMT) applications that work well at depth ranges that are typically out of reach for ERT.

A number of recent numerical studies have focused on the potential of using controlled source electromagnetics (CSEM) to monitor hydrocarbon reservoirs. Orange et al. (2009) and Lien and Mannseth (2008) considered CSEM monitoring for marine applications, while Wirianto et al. (2010) presented a feasibility study of land-based CSEM. They all conclude that CSEM monitoring is feasible, though not easy due to the diffusive character of EM signals and the low frequencies required to reach the reservoirs. For successful applications, this implies rather strong temporal contrasts and significant volumes experiencing temporally varying subsurface conditions.

The situation is more favorable for groundwater applications, as the targets of interest are typically located at shallower depths than hydrocarbon reservoirs. In a rare case study, Falgàs et al. (2009) successfully used AMT to monitor saltwater intrusion dynamics in a coastal aquifer in Spain. However, as the repeat surveys were not taken at the same positions, a time-lapse strategy could not be used and they inverted for independent models at each time. Nix (2005) monitored the spreading of a conductive tracer using scalar RMT data. By performing independent inversions along the same profile location at different times, models were obtained that were in fair accordance with groundwater data. There are to our knowledge no published studies concerning time-lapse inversions of AMT and RMT data despite the potential these techniques have for monitoring purposes.

One of the most widely used inversion strategies for geophysical inversion is minimum structure inversion, in which the model with the least structure is sought under the constraint that the model is consistent with the data and the estimated data errors (e.g., Constable et al., 1987; deGroot Hedlin and Constable, 1990; Siripunvaraporn and Egbert, 2000). To quantify model structure, the $l_2$-norm is commonly used. This is because its minimization results in a linear system to be solved, but it has the disadvantage that the models obtained are unrealistically smooth for many types of applications (Ellis and Oldenburg, 1994). Iteratively reweighted least squares (IRLS) algorithms make it possible to use non $l_2$-norms, while still solving a linear system at each iteration step. With such strategies, it is possible to obtain models with overall uniform regions separated by sharper interfaces. Last and Kubik (1983) used an IRLS scheme to minimize the total cross-sectional area of anomalous bodies when inverting 2D gravity data. Portniaguine and Zhdanov (1999) inverted 3D magnetic and gravity data by minimizing the volume in which the gradient of the properties is nonzero. Farquharson and Oldenburg (1998) minimized an $l_1$-type measure of the horizontal and vertical derivatives in the 2D inversion of electrical resistance data. Farquharson (2008) minimized an approximate $l_1$-norm of a combination of horizontal and vertical model differences together with differences between diagonal cells to better image dipping structures when inverting gravity and magnetotelluric (MT) data. Pilkington (2009) used the Cauchy norm to obtain sparse 3D magnetic models. The IRLS scheme has been successfully applied for different types of geophysical data to obtain compact models, but has only rarely been used in time-lapse applications (Ajo-Franklin et al., 2007).

The primary motivation of this paper is to evaluate, through numerical examples, to what extent inversion results based on AMT and RMT monitoring data can be enhanced by (1) time-lapse difference inversion; (2) incorporation of statistical information about the expected model updates; (3) using appropriate model norms to quantify temporal changes, (4) constraining model updates to predefined ranges. After presenting the theoretical background (section 2), we present the results of two numerical case studies (section 3). We then discuss the implications of these results for field-based applications (section 4) before making our conclusions (section 5).

## 2. Method
### 2.1. Basic magnetotelluric theory

Using distant source signals, the MT method measures the relations between frequency-dependent electric and magnetic field components that are sensitive to the resistivity structure of the Earth. Under the assumption of far field conditions, these fields are related through the impedance tensor **Z** (Cantwell 1960):

$$\begin{bmatrix} E_x(\omega) \\ E_y(\omega) \end{bmatrix} = \begin{bmatrix} Z_{xx}(\omega) & Z_{xy}(\omega) \\ Z_{yx}(\omega) & Z_{yy}(\omega) \end{bmatrix} \begin{bmatrix} H_x(\omega) \\ H_y(\omega) \end{bmatrix}, \quad (1)$$

where $\mathbf{E}(\omega) = [E_x(\omega), E_y(\omega)]^T$ is the horizontal electric field and $\mathbf{H}_h(\omega) = [H_x(\omega), H_y(\omega)]^T$ the horizontal magnetic field at a given angular frequency $\omega$, with T denoting transposition. The apparent resistivities $\rho_{ij}^{app}(\omega)$ and impedance phases $\varphi_{ij}(\omega)$ can be obtained from the impedance components, for example, for $Z_{xy}$:

$$\rho_{xy}^{app}(\omega) = \frac{1}{\omega \mu_0} \left| Z_{xy}(\omega) \right|^2, \quad (2)$$

$$\varphi_{xy}(\omega) = \tan^{-1}\left( \frac{\mathrm{Im}\, Z_{xy}(\omega)}{\mathrm{Re}\, Z_{xy}(\omega)} \right), \quad (3)$$

where $\mu_0 = 4\pi \times 10^{-7}$ Hm$^{-1}$ is the magnetic permeability of free space (analogous definitions hold for $Z_{yx}$). The geomagnetic transfer function (so-called tipper pointer) **T** relates the vertical and horizontal magnetic fields as

$$[H_z(\omega)] = [A(\omega) \quad B(\omega)] \begin{bmatrix} H_x(\omega) \\ H_y(\omega) \end{bmatrix} = \mathbf{T}^{\mathrm{T}} \begin{bmatrix} H_x(\omega) \\ H_y(\omega) \end{bmatrix}. \quad (4)$$

When considering 2D structures, Maxwell's equations can, for an appropriate rotation of the coordinate system, be decoupled into two independent modes: transverse-electric (TE) and transverse-magnetic (TM) (e.g., Zhang et al., 1987). Current flows parallel to the strike direction in the TE mode and perpendicular to it in the TM mode.

The RMT and AMT methods considered here differ from the MT technique in terms of the higher frequency range of the measurements and in terms of the origin of the sources used, but the governing equations generally remain the same. Classical MT modeling neglects the influence of displacement currents, but these must be included when considering high RMT frequencies acquired over very resistive formations (e.g., Linde and Pedersen, 2004b; Kalscheuer and Pedersen, 2007).

*2.2. Discrete deterministic inversion*

The inverse problem of deriving the multi-dimensional resistivity structure of the subsurface using impedance tensors and tipper pointers only, is both non-linear and underdetermined when considering finely discretized models. These challenges are most often addressed by using iterative methods based on successive linearization and by incorporating regularization constraints that strongly penalize model structure that deviates from a preconceived morphology. A number of inversion algorithms are available that are based on different numerical approaches (e.g. deGroot-Hedlin and Constable, 1990; Siripunvaraporn and Egbert, 2000; Rodi and Mackie, 2001).

Solutions based on smoothness-constrained least-squares formulations are often referred to as Occam inversion (Constable et al., 1987) and aim at finding the smoothest model that can explain the observed data within the assumed data errors. Given $N$ observed data $\mathbf{d}^{obs} = [d_1, d_2, ..., d_N]^{\mathrm{T}}$ and $M$ resistivity blocks $\mathbf{m} = [m_1, m_2, ..., m_M]^{\mathrm{T}}$ of constant properties with typically $M > N$, the inverse problem can in the 2D case be solved by minimizing the functional

$$W_\lambda(\mathbf{m}) = \alpha_y \left\| \partial_y (\mathbf{m} - \mathbf{m}_{ref}) \right\|_2^2 + \alpha_z \left\| \partial_z (\mathbf{m} - \mathbf{m}_{ref}) \right\|_2^2 + \lambda^{-1} \left\{ \left\| \mathbf{C}_\mathbf{d}^{-0.5} (\mathbf{d}^{obs} - \mathbf{F}[\mathbf{m}]) \right\|_2^2 - \chi_*^2 \right\}, \quad (5)$$

where $y$ and $z$ denote the horizontal and vertical directions of the 2D profile, respectively, $\alpha_i$, $i = y,z$ is the desired weight of smoothing in each direction, $\partial_i$, $i = y,z$ is the difference operator, $\mathbf{m}_{ref}$ is a reference model, $\mathbf{C}_\mathbf{d}^{-0.5} = \text{diag}\{\sigma_1^{-1},...,\sigma_N^{-1}\}$ aims at weighting the data with respect to their quality, $\mathbf{F}[\mathbf{m}]$ is the forward response of $\mathbf{m}$, $\chi_\mathbf{d}^2 = \left\| \mathbf{C}_\mathbf{d}^{-0.5}(\mathbf{d}^{obs} - \mathbf{F}[\mathbf{m}]) \right\|_2^2$ is the data misfit, $\chi_*^2$ is the desired data misfit and $\lambda$ is a trade-off parameter defining the weight in given to minimizing the model roughness. The desired model is found by iteratively solving:

$$\mathbf{m}_{k+1}(\lambda) = \left[ \left( \mathbf{C}_\mathbf{d}^{-0.5} \mathbf{J}_k \right)^T \mathbf{C}_\mathbf{d}^{-0.5} \mathbf{J}_k + \lambda \left( \alpha_y \partial_y^T \partial_y + \alpha_z \partial_z^T \partial_z \right) \right]^{-1} \left( \mathbf{C}_\mathbf{d}^{-0.5} \mathbf{J}_k \right)^T \mathbf{C}_\mathbf{d}^{-0.5} \hat{\mathbf{d}}_k + \mathbf{m}_{ref}, \quad (6)$$

where $\hat{\mathbf{d}}_k = \mathbf{d}^{obs} - \mathbf{F}[\mathbf{m}_k] + \mathbf{J}_k \Delta \mathbf{m}_k$, $\Delta \mathbf{m}_k = \mathbf{m}_k - \mathbf{m}_{ref}$, the subscripts $k$ and $k+1$ denote the previous and present iterations, $\mathbf{J}$ is the sensitivity matrix or Jacobian matrix, and $\lambda$ is determined through a line search. In the first iterations, $\lambda$ is chosen to minimize $\chi_\mathbf{d}^2$. When $\chi_\mathbf{d}^2 \leq \chi_*^2$, $\lambda$ is maximized under the constraint of satisfying $\chi_\mathbf{d}^2 \leq \chi_*^2$. The inverse of the matrix in square brackets in Eq. (6) is referred to as the generalized inverse.

The data misfit of inversion models is often represented in terms of the root mean square (RMS) misfit:

$$RMS = \sqrt{\frac{1}{N} \sum_{n=1}^{N} w_n^2}, \quad (7)$$

where $\mathbf{w} = \mathbf{C}_\mathbf{d}^{-0.5}(\mathbf{d}^{obs} - \mathbf{F}[\mathbf{m}])$. Given a Gaussian distribution of errors, the expected value of $\chi^2$ is $N$, which corresponds to an RMS misfit of 1.

Using the $l_2$-norm to quantify model structure, as in Occam inversion, favors smooth transitions of model properties over a number of model cells (e.g., Farquharson, 2008). If sharp transitions between geological units or anomalies with small spatial supports are expected, it is necessary to work with other model norms to obtain models in agreement with such pre-supposed properties. One numerically efficient way to do this is through iteratively reweighted least squares (IRLS) algorithms (e.g., Farquharson and Oldenburg, 1998; Portiaguine and Zhdanov, 1999; Ajo-Franklin et al., 2007; Pilkington, 2009). These algorithms rely on a least-square formulation similar to Eq. (6), but with the difference that reweighting matrices are defined after each iteration to approximate a given norm. This results in algorithms with similarly fast convergence characteristics as gradient-based formulations, but they allow resolving sharp interfaces or compact anomalies. The update to the IRLS solution of a non-linear inverse problem can be generalized as (e.g., Farquharson and Oldenburg, 1998; Menke, 1989; Siripunvaraporn and Egbert, 2000):

$$\mathbf{m}_{k+1}(\lambda) = \left[ \mathbf{J}_k^T (\mathbf{C}_\mathbf{d}^{-0.5})^T \mathbf{R}_{\mathbf{d},k} \mathbf{C}_\mathbf{d}^{-0.5} \mathbf{J}_k + \lambda (\mathbf{C}_\mathbf{m}^{-0.5})^T \mathbf{R}_{\mathbf{m},k} \mathbf{C}_\mathbf{m}^{-0.5} \right]^{-1} \mathbf{J}_k^T (\mathbf{C}_\mathbf{d}^{-0.5})^T \mathbf{R}_{\mathbf{d},k} \mathbf{C}_\mathbf{d}^{-0.5} \hat{\mathbf{d}}_k + \mathbf{m}_{ref}, \quad (8)$$

where $\mathbf{C}_\mathbf{m}^{-0.5}$ indicates a more general and flexible model regularization matrix than the difference operator in Eq. (5), and $\mathbf{R}_i$, $i=\mathbf{m},\mathbf{d}$ is a reweighting matrix that is recalculated after each iteration and that depends on the norm chosen.

In this work, the amount of model structure is quantified by considering a given norm of the vector $\mathbf{x} = \mathbf{C}_\mathbf{m}^{-0.5} \Delta \mathbf{m}$. There are several norms that can be used to emphasize different aspects of model structure. For example, Ekblom's perturbed $l_p$-norm

$$\phi(\mathbf{x}) = \sum_{m=1}^{M} \left( x_m^2 + \gamma^2 \right)^{p/2}, \quad (9)$$

where $\gamma$ is a small number with respect to $x_m$. Choosing $p = 1$ makes it possible to approximate the $l_1$-norm with the advantage that its derivative exists at $x = 0$. The $l_1$-norm imposes penalizations proportional to the values of $\mathbf{x}$, contrary to the $l_2$-norm, which provides an enhanced penalization of large values. Ekblom's norm can be implemented in the IRLS algorithm by taking $\mathbf{R}$ as:

$$R_{ii} = p(x_i^2 + \gamma^2)^{p/2-1}. \quad (10)$$

Last and Kubik (1983) and Portniaguine and Zhdanov (1999) use a minimum support measure defined as

$$\phi(\mathbf{x}) = \sum_{m=1}^{M} \frac{x_m^2}{\left(x_m^2 + \gamma^2\right)}, \tag{11}$$

with a corresponding **R**:

$$R_{ii} = \frac{2\gamma^2}{(x_i^2 + \gamma^2)^2}. \tag{12}$$

This norm is proportional to the number of nonzero elements of **x**. The Cauchy norm

$$\phi(\mathbf{x}) = \sum_{m=1}^{M} \ln\left(1 + \frac{x_m^2}{\gamma^2}\right), \tag{13}$$

with

$$R_{ii} = \frac{1}{(x_i^2 + \gamma^2)}, \tag{14}$$

is another norm used to obtain an **x** with few values different from zero (Sacchi and Ulrych, 1996; Pilkington, 2009). The norm decreases as more elements of $x_m$ are smaller than $\gamma$. The choice of $\gamma$ controls the amplitudes and the fractions of non-zero values.

For the numerical experiments considered in this work, we have found that taking $\gamma = \sum_{m=1}^{M}\left(\frac{x_m}{M}\right)$ leads to satisfactory solutions for all cases considered. Schemes based on IRLS must allow for several model iterations before the final model is computed, such that the reweighting of the regularization term is consistent with the final model. As shown in Eq. (8), the IRLS scheme can also be used to apply different norms to quantify data misfit. In these numerical investigations, we assume and impose a Gaussian distribution of the data residuals. The data misfit was consequently quantified with a classical $l_2$-norm (c.f., eq. (7)), which is optimal for Gaussian errors.

*2.3. Time-lapse inversion*

When monitoring temporal changes in subsurface properties it is advantageous to leave the sensors in place during the monitoring period. The errors in the resulting time-lapse data acquired for the same sensor configuration are likely to share a repeatable systematic component, which can be largely removed in the time-lapse inversion. Following LaBrecque and Yang (2001), the observed data at time-lapse *t* can be expressed as:

$$\mathbf{d}_t^{obs} = \mathbf{F}[\mathbf{m}_t] + \boldsymbol{\varepsilon}_{sys} + \boldsymbol{\varepsilon}_{r,t}, \tag{15}$$

where $\boldsymbol{\varepsilon}_{r,t}$ is a random observational error that is varying in time and is thus different for each data set and $\boldsymbol{\varepsilon}_{sys}$ is a systematic contribution that is present at all times. Systematic errors can be related to modeling errors, bias introduced by ground coupling problems or improperly calibrated sensors, deviations from 2D assumptions or geometrical errors (e.g., incorrect electrode positioning or profiles that are not perfectly aligned).

Time-lapse inversion algorithms can be defined in different ways but aim generally at removing the systematic contribution to allow resolving minute changes in subsurface properties over time. In a first step, the model at $t = 0$, $\mathbf{m}_0$, is obtained by means of a standard inversion (see section 2.2.) using the data acquired before any perturbation to the system. Next, the residuals $\mathbf{r}_0 = \mathbf{d}_0^{obs} - \mathbf{F}[\mathbf{m}_0] = \varepsilon_{sys} + \varepsilon_{r,0}$, are removed from the data acquired at all subsequent times:

$$\tilde{\mathbf{d}}_t^{obs} = \mathbf{d}_t^{obs} - \mathbf{r}_0 = \mathbf{F}[\mathbf{m}_t] + \varepsilon_{r,t} - \varepsilon_{r,0}. \tag{16}$$

Since the systematic component has been removed by differencing, the new corrected data sets have the advantage of being less error contaminated, provided the common situation concerning the standard deviations of the different error sources that $\sigma_{sys} > \sqrt{\sigma_{r,0}^2 + \sigma_{r,t}^2}$ (e.g., Doetsch et al., 2010). Furthermore, $\mathbf{m}_0$ can be used as the reference model $\mathbf{m}_{ref}$ for the following inversions, so that the model regularization is applied to the model update with respect to the model reference.

*2.4. Stochastic regularization*

Statistical information of the expected model update with respect to $\mathbf{m}_{ref}$ can be used to constrain time-lapse inversions and thereby accurately include statistical properties as a priori information. Maurer et al. (1998) showed that regularization based on model covariance models, so-called stochastic regularizations, uniquely define the relative contribution of penalizing roughness (i.e., to obtain smooth models) with respect to damping (i.e., to obtain models that are close to $\mathbf{m}_{ref}$). A given covariance function is used to compute the model covariance matrix, which is then inverted to obtain the regularization matrix $\mathbf{C}_\mathbf{m}^{-0.5}$. When the correlation function is stationary throughout a uniform grid, the covariance matrix can be inverted efficiently through circulant embedding and using the diagonalization theorem of circulant matrices (Dietrich and Newsam, 1997; Linde et al., 2006).

Here we consider the exponential correlation function for a two-dimensional domain that is defined as

$$r(l) = c \exp(-l), \qquad (17)$$

where $c$ is the variance and $l$ is

$$l = \sqrt{\left(\frac{h_y}{I_y}\right)^2 + \left(\frac{h_z}{I_z}\right)^2}, \qquad (18)$$

where $h_i$, $i = y,z$ is the separation between two points and $I_i$, $i = y,z$ is the integral scale that characterizes the spatial correlation in each direction (e.g., Rubin, 2003). When the distance between two points equals the integer scale, their correlation is $1/e \approx 37\%$. As the same constant $c$ is assumed for all model parameters, the multiplication of $\mathbf{C}_\mathbf{m}^{-0.5}$ in Eq. (8) results in the inverse of the constant $c$ being multiplied with $\lambda$. As we perform a line search for $\lambda$, we set $c = 1$.

*2.5. Guiding the model update*

Constraints regarding model parameter updates can be incorporated through Lagrange multipliers (e.g., Menke, 1989, Chapter 3.10). If the model values are only expected to decrease with respect to $\mathbf{m}_{ref}$ (e.g., electrical resistivity is expected to decrease due to the application of a saline tracer), one can guide the inversion by penalizing model parameters exhibiting positive deviations from $\mathbf{m}_{ref}$ in iteration $k$, to have values closer to $\mathbf{m}_{ref}$ in iteration $k+1$ by adding the constraint $\mathbf{H}_k \Delta \mathbf{m}_{k+1} = \mathbf{0}$, where $\mathbf{H}_k$ is of size $M_\mathbf{v} \times M$, and $M_\mathbf{v}$ is the number of elements of $\Delta\mathbf{m}_k$ that are positive and should be guided to be zero. In iteration $k+1$, $\mathbf{H}_k$ is generated from $\Delta\mathbf{m}_k$ as

$$H_k^{ij} = \begin{cases} 1, & \text{if } \Delta m_k^j > 0 \text{ is the } i\text{-th value to penalize} \\ 0, & \text{otherwise} \end{cases}$$

The constraint equation must be solved simultaneously with Eq. (8) (Menke, 1989), which results in an augmented system of equations:

$$\begin{bmatrix} \mathbf{m}_{k+1}(\lambda) \\ \mathbf{v}(\lambda) \end{bmatrix} = \begin{bmatrix} \mathbf{J}_k^T (\mathbf{C}_\mathbf{d}^{-0.5})^T \mathbf{R}_{\mathbf{d},k} \mathbf{C}_\mathbf{d}^{-0.5} \mathbf{J}_k + \lambda (\mathbf{C}_\mathbf{m}^{-0.5})^T \mathbf{R}_{\mathbf{m},k} \mathbf{C}_\mathbf{m}^{-0.5} & \mathbf{H}_k^T \\ \mathbf{H}_k & \mathbf{0} \end{bmatrix}^{-1} \begin{bmatrix} \mathbf{J}_k^T (\mathbf{C}_\mathbf{d}^{-0.5})^T \mathbf{R}_{\mathbf{d},k} \mathbf{C}_\mathbf{d}^{-0.5} \hat{\tilde{\mathbf{d}}}_k \\ \mathbf{0} \end{bmatrix} + \begin{bmatrix} \mathbf{m}_{\text{ref}} \\ \mathbf{0} \end{bmatrix} \quad (19)$$

where $\hat{\tilde{\mathbf{d}}}_k = \tilde{\mathbf{d}}_t^{\text{obs}} - \mathbf{F}[\mathbf{m}_k] + \mathbf{J}_k \Delta\mathbf{m}_k$, and we solve for the $M$ new model parameters in vector $\mathbf{m}_{k+1}$, plus the $M_\mathbf{v}$ unknown Lagrange multipliers in vector $\mathbf{v}$.

## 3. Numerical examples

### 3.1. A shallow prism

A very simple synthetic test case was considered to investigate the influence of the different regularizations and norms presented in section 2 on the time-lapse inversions results. The test case consists of a model that changes between time instance $t = 0$ (Fig. 1a) and $t = 1$ (Fig. 1b), in which the only difference is a conductive prism with a cross-sectional area of $6\times6$ m$^2$ that appears at $t = 1$.

The forward responses of the models were computed for both the TE and TM modes including the real and imaginary parts of the tipper pointer, at 7 stations with a separation of 5 m. A total of 10 frequencies regularly spaced in logarithmic scale (two frequencies per octave) were used in the RMT frequency range of 10 to 226 kHz, which resulted in 420 data points. A mesh of $58\times104$ cells was used for the forward computations and the inversions, including 10 rows needed to model the air. The central part of the mesh, which is shown in Figs. 1 and 2, has a $1\times1$ m$^2$ discretization. All the forward calculations and inversions presented here were calculated using a modified version (cf. Kalscheuer et al., 2010) of the REBOCC code (Siripunvaraporn and Egbert, 2000).

To simulate the time-lapse data, two types of errors were added to the forward responses of the synthetic models (see section 2.3.). Uncorrelated Gaussian noise with zero mean was considered in all cases. For the impedances, standards deviations of $\sigma_{sys} = 10\%$ and $\sigma_{r,0} = \sigma_{r,1} = 2\%$ were considered, whereas for the real and imaginary part of the tipper $\sigma_{sys} = 0.02$ and $\sigma_{r,0} = \sigma_{r,1} = 0.005$ were used. Inversions of the synthetic data were performed using the different inversion and regularization schemes defined in sections 2.2-2.3. To allow for more iterations before convergence of the IRLS inversions, a maximum of five points were evaluated to determine $\lambda$ and thereby decrease the convergence rate.

A standard smoothness-constrained least squares inversion, referred to as Occam in the following (see Eq. (6)), was used to obtain independently inverted models at each time. The errors assumed were 10.2% (i.e., $\sqrt{10^2 + 2^2}$) for the impedance elements and 0.0206 (i.e., $\sqrt{0.02^2 + 0.005^2}$) for the tipper components, and a half-space of 100 $\Omega$m was used as the starting model. For the time-lapse inversions, new time-lapse corrected data sets were created using Eq. (16). The reference model used was the one obtained with an Occam inversion for $t = 0$ (Fig. 1c). The errors assumed were corresponding to the random components of the noise added.

Figure 2 shows the true model update (Fig. 2a) together with the model updates obtained with the different inversion schemes (Figs. 2b-h). All the models in Fig. 2 fit the data with RMS ≤ 1.05. Figure 2b shows the model update obtained by differencing the two separate Occam inversions. The region of change is approximately detected, but it is resolved as a smoothly varying feature of strongly overestimated extent that is centered below the actual anomaly. Furthermore, positive updates representing artifacts appear on the sides of the model. Figure 2c shows the result of applying the time-lapse inversion to the traditional Occam inversion (i.e., time-lapse corrected data, but with smoothness constraints using an $l_2$ measure). The lower error-level in the time-lapse corrected data helps to better constrain the geometry of the model update, which is considerably more focused than the previous example. However, the model update is still rather smooth due to the $l_2$ measure of model structure and oscillations representing inversion artifacts are still visible. Figure 2d shows the model update obtained using stochastic regularization with the $l_2$-norm in the time-lapse scheme. Assuming that points more than 6 m apart are poorly correlated, the integral scales were chosen as $I_y = I_z = 3$m (correlation is less than 14% for separations larger than two integral scales, see Sec. 2.4). The model obtained is very similar to that of Fig. 2c. Figures 2e-g show the results of applying the stochastic regularization to the time-lapse inversion using the perturbed $l_1$-norm, minimum support norm and Cauchy norm as measures of model structure, respectively. The delineations of the anomalous region are much sharper and the oscillations shown in Figs. 2b-d have essentially been removed. Some cells with positive resistivity changes can be seen in the three models, close to the receiver stations in Figs. 2e and 2g and inside the prism in Fig. 2f. These features disappear when penalizing positive changes to the model through Lagrange multipliers (see Eq. (19)). The model update obtained with the penalized inversion using the $l_1$-norm is shown in Fig. 2h. Similar results were found with the Cauchy and minimum support measures. A feature that is common to all the cases where non $l_2$-norms were used is an overestimation of the inferred magnitude of the model update in the center of the prism.

Figure 3 shows horizontal slices at a depth of 8 m through the models in Fig. 2. The sharp changes and large amplitudes obtained with the non $l_2$-norms contrast strongly with the smoothly oscillating updates obtained with the $l_2$-norm inversions. The rather small amplitude observed in the curve representing the Cauchy norm model (Fig. 2g) is due to the maximum value not being found at 8 m, but at a depth of 7 m.

We define two measures to quantify the similarity between the proposed model updates and the true model update. The first one is an average of the final **Δm** for the cells located in the region where the true prism is located. Since inversions of electromagnetic data are essentially always working with logarithms of resistivity (or conductivity), the same units were used to compute the average. The second measure is an average of the amplitudes of **Δm** outside the region of the prism, which is zero for the true difference. It quantifies how much structure a certain solution is adding outside the region where the true changes occur.

The values of the two measures together with the RMS for the different inversion cases are given in Table 1. For the true model difference, the means inside and outside the anomaly are -1 and 0, respectively. The traditional Occam scheme has a mean of -0.27 inside the anomaly and 0.078 in the outside region. For the time-lapse cases, the $l_2$-norm gives better estimates of the average magnitude of the update, with means of -0.54 and -0.58, but puts a lot of structure outside the region of changes (means of 0.058 and 0.053). On the other hand, the non $l_2$-norms have a smaller average update, but the structure outside the true anomaly has decreased with 1 or 2 orders of magnitude. When the update calculated using the Cauchy norm is constrained to be negative, the average value within the prism is closer to the actual value (-0.53).

*3.2. Seawater intrusion example*

The second more complex test case is inspired by the experiment of Falgàs et al. (2009), who monitored a seawater-freshwater mixing zone over time using AMT. The models at $t = 0$ and $t = 1$ are shown in Figs. 4a and 4b. The models comprise a 100 m thick 100 Ωm layer, in which seawater intrusion occurs in the lower 50 m. The aquifer overlies a 630 Ωm half-space. The sea is modeled in the rightmost upper corner with a resistivity of 0.3 Ωm. The seawater encroachment is represented with a linearly increasing resistivity, from 3 Ωm corresponding to a rock completely saturated with seawater to 100 Ωm corresponding to freshwater conditions. At $t = 1$, the seawater-freshwater interface has advanced 300 m inland with respect to the situation at $t = 0$.

The forward responses were simulated considering 10 stations with a spacing of 160 m and a frequency range of 10 Hz to 116 kHz (two frequencies per octave as in the prism example). Within the region of interest, the cell size is 20×10 m². The simulated TE mode, TM mode and tipper data were noise contaminated with $\sigma_{sys} = 5\%$ and $\sigma_{r,0} = \sigma_{r,1} = 2\%$ for the impedances, and $\sigma_{sys} = 0.02$ and $\sigma_{r,0} = \sigma_{r,1} = 0.005$ for the tipper components.

Figure 4c shows the result of Occam inversion using the noise-contaminated data. A half-space of 100 Ωm was used as the starting model and a five times larger regularization weight was applied in the horizontal direction to emphasize the layered nature of the model. The model obtained is similar to the one shown by Falgàs et al. (2009). The seawater encroachment is clearly detected and the upper part of the aquifer is well resolved, but the lower part is imaged with a gradual increase in the resistivity rather than a sharp transition.

Figure 5a shows the difference between the true models at times $t = 1$ and $t = 0$. Changes in the horizontal direction are smooth, while the transition between layers in the vertical direction is sharp. The model difference between the two independent Occam inversions is shown in Fig. 5b. The upper interface of the time-lapse anomaly is well resolved, whereas the lower interface is very diffuse and extends to large depths. The lateral extension of the anomalous region is well resolved, but a positive artifact is shown to the right. The model update given by the time-lapse inversion with stochastic regularization ($l_y = 200$ m and $l_z = 20$ m) using the $l_2$-norm (Fig. 5c) and the Occam inversion as reference model better defines the lower interface, but presents more regions of positive inversion artifacts. When the perturbed $l_1$-norm is used (Fig. 5d), the positive changes observed in the lower part of the profile disappear and the transitions get sharper, but the positive changes towards the seaside prevail. Figures 5e and 5f show the model updates obtained when penalizing positive values using the perturbed $l_1$-norm and Cauchy norm, respectively. The time-lapse target is well resolved and no oscillations are observed outside the region of the time-lapse target. Furthermore, in the case of the perturbed $l_1$-norm, the smooth horizontal transition is respected and the lateral extent of the anomaly corresponds overall quite well with the time-lapse target. This is not the case for the Cauchy norm, which resolves the time-lapse change as being laterally more compact than it really is. Furthermore, some cells with positive resistivity updates can still be observed for this model. Inversions using the minimum support norm with and without penalizing positive changes did not converge for this model.

Horizontal and vertical cuts of the models shown in Fig. 5 are presented in Fig. 6a and 6b, respectively. The true difference (shown in black in Fig. 6a) is smoothly varying in the horizontal direction. All the inversion schemes reproduce this transition rather well, except for the constrained Cauchy norm, shown in blue, which presents very small updates at this depth of 80 m. The largest differences between the models can be seen in the right part of the figure, that is, the region closer to the sea. Artifacts are present when using both the $l_2$-norm and the non-constrained $l_1$-norm. These artifacts disappear only when penalizing the positive updates, which results in a curve that closely follows the true model update at all points in the case of the perturbed $l_1$-norm. In the vertical cut (Fig. 6b), the results are similar to those of the prism example: the time-lapse inversions better constrain the model update, the non $l_2$-norms make the transitions sharper and the negativity constraints reduce or completely eliminate positive value updates.

Table 2 shows the comparison statistics for each model in Fig. 5. The average value of the model update is -0.65 inside the true anomaly and 0 outside. Inside the anomaly, the mean magnitude is well estimated in all cases except for the Cauchy norm, which presents some positive updates inside the region of true change. Outside the anomaly, Occam inversion is again the method that puts the most structure (0.103) (c.f. Table 1). The time-lapse inversion using the $l_2$-norm has a mean of absolute values of 0.098, and the perturbed $l_1$-norm 0.072. Only when the negativity constraints are added, the mean of the absolute values outside the anomalous region is reduced by one order of magnitude.

## 4. Discussion

Falgàs et al. (2009) demonstrated convincingly that AMT monitoring allows resolving seasonal seawater-freshwater dynamics. The aim of this work was to investigate through numerical examples to what extent these types of results could be further improved by using more refined inverse formulations.

As expected, removing errors that are constant over time ($\varepsilon_{sys}$) clearly yield improved models for the two case studies. The model updates provided by differencing independent Occam inversions (Figs. 2b and 5b) were unnecessarily diffuse compared with a difference inversion that otherwise is based on the same type of objective function (i.e., smoothness constraints and a $l_2$-norm). This improvement is explained by error cancellation in the difference inversion scheme as outlined by LaBrecque and Yang (2001).

Compared with smoothness constraints, stochastic regularization offers added flexibility in imposing statistical information of the expected model morphology (e.g., Linde et al., 2006). In our case, this was used to add information about the expected scale of temporal changes in the model (Doetsch et al., 2010). No specific integer scale can be defined for the model shown in Fig. 1b because of the superposition of geological layers and the time-lapse anomaly, while this is easier when inverting for the model update (Fig. 2a). For the examples considered in this study, we do not find any significant differences between the time-lapse inversion results based on an $l_2$-norm when using stochastic regularization (Fig. 1d) compared with smoothness constraints (Fig. 1c). In fact, both types of models are unsuitable as they are overly smooth and display oscillations in the region around the true anomaly.

To obtain sharper transitions, we applied non $l_2$-norms in a similar manner as Farquharson and Oldenburg (1998), Portiaguine and Zhdanov (1999), and Pilkington (2009), but to time-lapse data (Ajo-Franklin et al., 2007). Using the perturbed $l_1$-norm, the Cauchy norm and the minimum support measures, we obtained compact model updates with a significant decrease in structure outside the true anomaly (Figs. 2e-g and 5d), but with the magnitude in some of the model cells being overestimated (Figs. 3 and 5).

The seawater on the right side in the saltwater intrusion example resulted in significant artifacts in the time-lapse inversions, especially for the non $l_2$-norms. Even if the values in this region at $t = 1$ were the same as at $t = 0$, large positive structures appeared for the three non-traditional norms used (only the perturbed $l_1$-norm example is shown in Fig. 5d). As we were considering a time at which the seawater-freshwater transition zone advances inland, it was natural to penalize positive changes in resistivity. Each cell with constraints adds a dimension to the matrix that has to be inverted, which can be computationally demanding in terms of memory and computing time. For the examples considered here, the number of elements of the model update that need to be penalized constitutes a significant percentage of the model blocks only in the first iterations. Note that penalizing positive values by adding Lagrange multipliers does not ensure that no positive cells are going to be found in the model update.

Indeed, a few positive cells can be observed in the model update calculated with the Cauchy norm when applying the negativity constraints.

Another technical issue is that the reweighting needed for the non-traditional norm increases the condition number of the generalized inverse and also tends to increase the non-linearity of the inverse problem. Of the three norms considered, the perturbed $l_1$-norm was found to be the most robust in the sense that it did not significantly change the condition number of the matrices to be inverted compared with the $l_2$-norm case.

## 5. Conclusions

We find that inversion results based on monitoring of uniform inducing field electromagnetic data (RMT and AMT in the examples considered) can be much improved by using difference inversion and by incorporating information regarding the expected changes in model properties over time. Compact and sharper model updates were obtained by combining stochastic regularization and non $l_2$-norms implemented through an IRLS procedure. In particular, the perturbed $l_1$-norm was found to be both robust and allowing for smooth variations, not creating compact models when this was not the case. Penalizing model updates with non-physical variations (e.g., increases in resistivity when saltwater is intruding) was shown to be successful not only in avoiding inversion artifacts, but also, in the case of the perturbed $l_1$-norm, to better determine the magnitudes of the time-lapse changes. A characteristic of all model updates computed with non $l_2$-norms is the overestimation of the magnitudes of the changes in some cells. Such overestimations can be removed using Lagrange multipliers, similarly as for the negativity constraints, given that the expected maximum amplitudes of the true changes are known or can be adequately assessed. The presented inversion methodology will in the future be applied to field data, which will require the development of robust transfer function estimation procedures of time-lapse data.

**Acknowledgments**

This research was supported by the Swiss National Science Foundation under grant 200021-130200. We thank Juanjo Ledo and an anonymous reviewer for their useful comments. The original EM inversion code EMILIA, and the presented implementations, are available upon request by contacting Thomas Kalscheuer: thomas@aug.ig.erdw.ethz.ch.

**References**

Ajo-Franklin, J.B., Minsley, B.J., Daley, T.M., 2007, Applying compactness constraints to differential traveltime tomography. Geophysics 72, R67-R75. doi:10.1190/1.2742496.

Beylich, A.A., Kolstrup, E., Linde, N., Pedersen, L.B., Thyrsted, T., Gintz, D., and Dynesius, L., 2003, Assessment of chemical denudation rates using hydrological measurements, water chemistry analysis and electromagnetic geophysical data. Permafrost and Periglacial Processes 14, 387-397. doi:10.1002/ppp.470.

Cantwell, T., 1960. Detection and analysis of low-frequency magnetotelluric signals. Ph.D. Thesis, Department of Geology and Geophysics, M.I.T., Cambridge.

Constable, S.C., Parker, R.L., Constable, C.G., 1987. Occam's inversion: A practical algorithm for generating smooth models from electromagnetic sounding data. Geophysics 52, 289–300.

deGroot-Hedlin, C.D., Constable, S.C., 1990. Occam's inversion to generate smooth, two-dimensional models from magnetotelluric data. Geophysics 55, 1613-1624.

Dietrich, C.R., Newsam, G.N., 1997. Fast and exact simulation of stationary Gaussian processes through circulant embedding of the covariance matrix. SIAM Journal of Scientific Computing 18, 1088-1107.

Doetsch, J., Linde, N., Binley, A., 2010. Structural joint inversion of time-lapse crosshole ERT and GPR traveltime data. Geophysical Research Letters 37, L24404. doi:10.1029/2010GL045482.

d'Ozouville, N., Auken, E., Sorensen, K., Violette, S, de Marsily, G., Deffontaines, B., Merlen, G., 2008. Extensive perched aquifer and structural implications revealed by 3D resistivity mapping in a Galapagos volcano. Earth and Planetary Science Letters 269, 517-521. doi: 10.1016/j.epsl.2008.03.011.

Ellis, R.G., Oldenburg, D.W., 1994. Applied geophysical inversion. Geophysical Journal International 116, 5–11. doi: 10.1111/j.1365-246X.1994.tb02122.x.

Falgàs, E., Ledo, J., Marcuello, A., Queralt, P., 2009. Monitoring freshwater-seawater interface dynamics with audiomagnetotelluric data. Near Surface Geophysics 7, 391-399.

Farquharson, C.G., 2008. Constructing piecewise-constant models in multidimensional minimum-structure inversions. Geophysics 73, K1-K9. doi: 10.1190/1.2816650.

Farquharson, C.G., Oldenburg, D.W., 1998. Nonlinear inversion using general measures of data misfit and model structure. Geophysical Journal International 134, 213-227.

Fitterman, D.V., and Stewart, M.T., 1986, Transient electromagnetic sounding for groundwater. Geophysics 51, 995-1005.

Kalscheuer, T., Pedersen, L.B., 2007. A non-linear truncated SVD variance and resolution analysis of two-dimensional magnetotelluric models. Geophysical Journal International 169, 435–447. doi:10.1111/j.1365-246X.2006.03320.x

Kalscheuer, T., García Juanatey, M., Meqbel, N., Pedersen, L.B., 2010. Non-linear model error and resolution properties from two-dimensional single and joint inversions of direct current resistivity and radiomagnetotelluric data. Geophysical Journal International 182, 1174–1188. doi: 10.1111/j.1365-246X.2010.04686.x.

Kemna, A., Vanderborght, J., Kulessa, B., Vereecken, H., 2002. Imaging and characterization of subsurface solute transport using electrical resistivity tomography (ERT) and equivalent transport models. Journal of Hydrology 267, 125-146. doi: 10.1016/S0022-1694(02)00145-2.

LaBrecque, D.J., Yang, X., 2001. Difference inversion of ERT data: a fast inversion method for 3-D in situ monitoring. Journal of Environmental & Engineering Geophysics 5, 83–90.

Last, B.J., Kubik, K., 1983. Compact gravity inversion, Geophysics 48, 713, doi:10.1190/1.1441501

Lien, M., Mannseth, T., 2008. Sensitivity study of marine CSEM data for reservoir production monitoring. Geophysics 73: F151-163.

Linde, N., Pedersen, L.B., 2004a. Evidence of electrical anisotropy in limestone formations using the RMT technique. Geophysics 69, 909-916. doi: 10.1190/1.1778234.

Linde, N., and Pedersen, L.B., 2004b. Characterization of a fractured granite using radio magnetotelluric (RMT) data. Geophysics, 69, 1155-1165. Doi:10.1190/1.1801933.

Linde, N., Binley, A., Tryggvason, A., Pedersen, L.B., Revil, A., 2006. Improved hydrogeophysical characterization using joint inversion of cross-hole electrical resistance and ground-penetrating radar traveltime data. Water Resources Research 42, W12404. doi:10.1029/2006WR005131.

Maurer, H., Holliger, K., Boerner D.E., 1998. Stochastic regularization: Smoothness or similarity? Geophysical Research Letters 25, 2889–2892. doi:10.1029/98GL02183.

Menke, W., 1989. Geophysical data analysis: Discrete inverse theory, Vol. 45 of International Geophysics Series, Academic Press, London.

Minsley, B.J., Ajo-Franklin, J., Mukhopadhyay, A., Morgan, F.D., 2011, Hydrogeophysical methods for analyzing aquifer storage and recovery systems. Ground Water 49, 250-269. doi: 10.1111/j.1745-6584.2010.00676.x.

Nix, B., 2005. Radiomagnetotellurik-messungen zur räumlichen und zeitlichen ausbreitung ines Grundwasser-Tracers. Diploma thesis, University of Cologne, Germany.

Orange, A., Key, K., Constable S., 2009. The feasibility of reservoir monitoring using time-lapse marine CSEM. Geophysics 74, F21–F29. doi:10.1190/1.3059600.

Pilkington, M., 2009. 3D magnetic data-space inversion with sparseness constraints. Geophysics 74: L7-15.

Portniaguine O., Zhdanov, M.S., 1999. Focusing geophysical inversion images: Geophysics 64, 3, 874-887.

Rodi W., Mackie R., 2001. Nonlinear conjugate gradients algorithm for 2-D magnetotelluric inversion. Geophysics 66, 174-187.

Rubin, Y. Hubbard, S.S. (Eds.), 2005. Hydrogeophysics. Springer.

Rubin, Y., 2003. Applied stochastic hydrogeology, Oxford University Press.

Sacchi, M.D., Ulrych, T.J, 1996. Estimation of the discrete Fourier transform: a linear inversion approach. Geophysics 61, 1128-1136.

Siripunvaraporn, W., Egbert, G., 2000. An efficient data-subspace inversion method for 2-D magnetotelluric data. Geophysics 65, 791-803.

Tezkan, B., 1999. A review of environmental applications of quasi-stationary electromagnetic techniques. Surveys in Geophysics 20, 279-308. doi: 10.1023/A:1006669218545.

Wirianto, M., Mulder, W. A., Slob, E. C., 2010. A feasibility study of land CSEM reservoir monitoring in a complex 3-D model. Geophysical Journal International 181, 741–755. doi: 10.1111/j.1365-246X.2010.04544.x

Zhang, P., Roberts, R.G., Pedersen, L.B., 1987. Magnetotelluric strike rules. Geophysics 52, 267-278.

**Table 1.** Statistics of performance measures for the model differences shown in Fig. 2.

| Inversion Strategy | Mean of **Δm** inside the true anomaly $\log_{10} \rho$ (Ωm) | Mean of \|**Δm**\| outside the true anomaly $\log_{10} \rho$ (Ωm) | RMS |
|---|---|---|---|
| True Model Difference | -1 | 0 | 0.97 |
| Difference of Occam's inversions | -0.27 | 0.078 | 0.97 |
| TL Occam's inversion | -0.54 | 0.058 | 1.00 |
| TL Stoch. Reg. $l_2$-norm | -0.58 | 0.053 | 0.98 |
| TL Stoch. Reg. Perturbed $l_1$-norm | -0.5 | 0.015 | 0.99 |
| TL Stoch. Reg. Minimum Support | -0.3 | 0.006 | 1.02 |
| TL Stoch. Reg. Cauchy Norm | -0.3 | 0.007 | 1 |
| TL Stoch. Reg. Cauchy Norm + negativity constraints | -0.53 | 0.014 | 1.04 |

**Table 2.** Statistics of performance measures for the model differences shown in Fig. 5.

| Inversion Strategy | Mean of **Δm** inside the true anomaly $\log_{10} \rho$ (Ωm) | Mean of \|**Δm**\| outside the true anomaly $\log_{10} \rho$ (Ωm) | RMS |
|---|---|---|---|
| True Model Difference | -0.65 | 0 | 0.96 |
| Difference of Occam's inversions | -0.62 | 0.103 | 1.01 |
| TL Stoch. Reg. $l_2$-norm | -0.67 | 0.098 | 1.04 |
| TL Stoch. Reg. Perturbed $l_1$-norm | -0.67 | 0.072 | 1.00 |
| TL Stoch. Reg. Pert. $l_1$-norm + negativity constraints | -0.68 | 0.033 | 1.05 |
| TL Stoch. Reg. Cauchy Norm + negativity constraints | -0.44 | 0.021 | 1.04 |

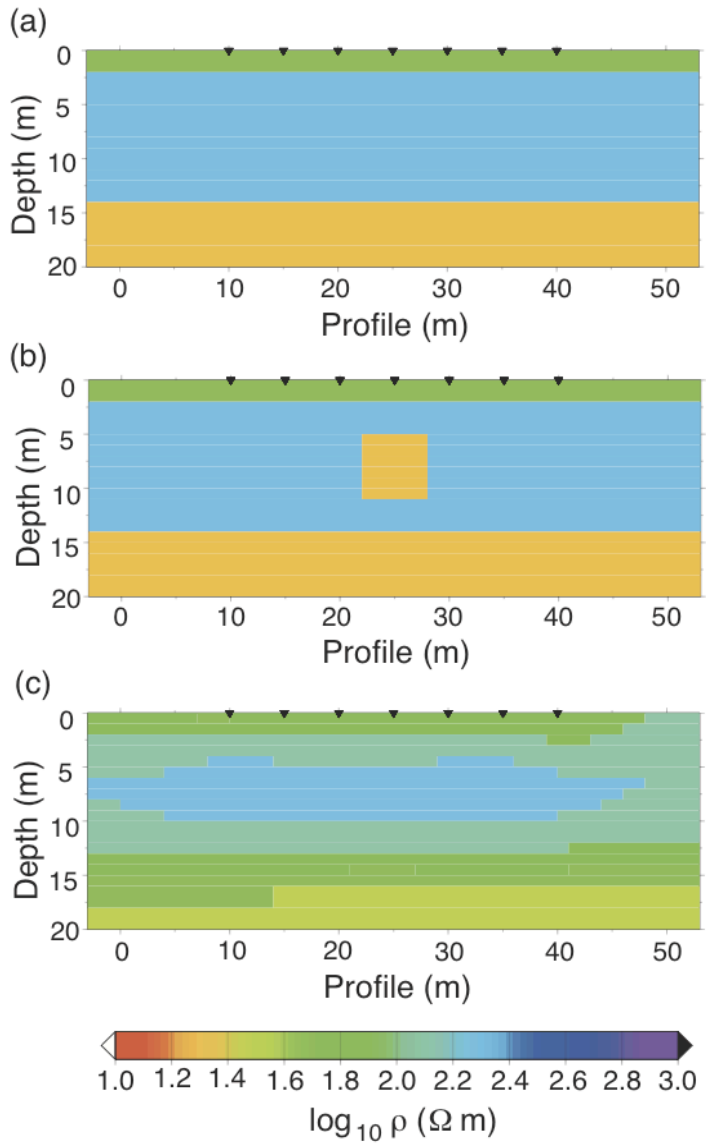

**Fig. 1**. Synthetic 2D models used to generate the data at (a) $t = 0$ and (b) $t = 1$ for the shallow prism model. (c) Reference model for the time-lapse inversions obtained by inverting the data at $t = 0$ with an Occam algorithm. The triangles at the top of the figures indicate station locations.

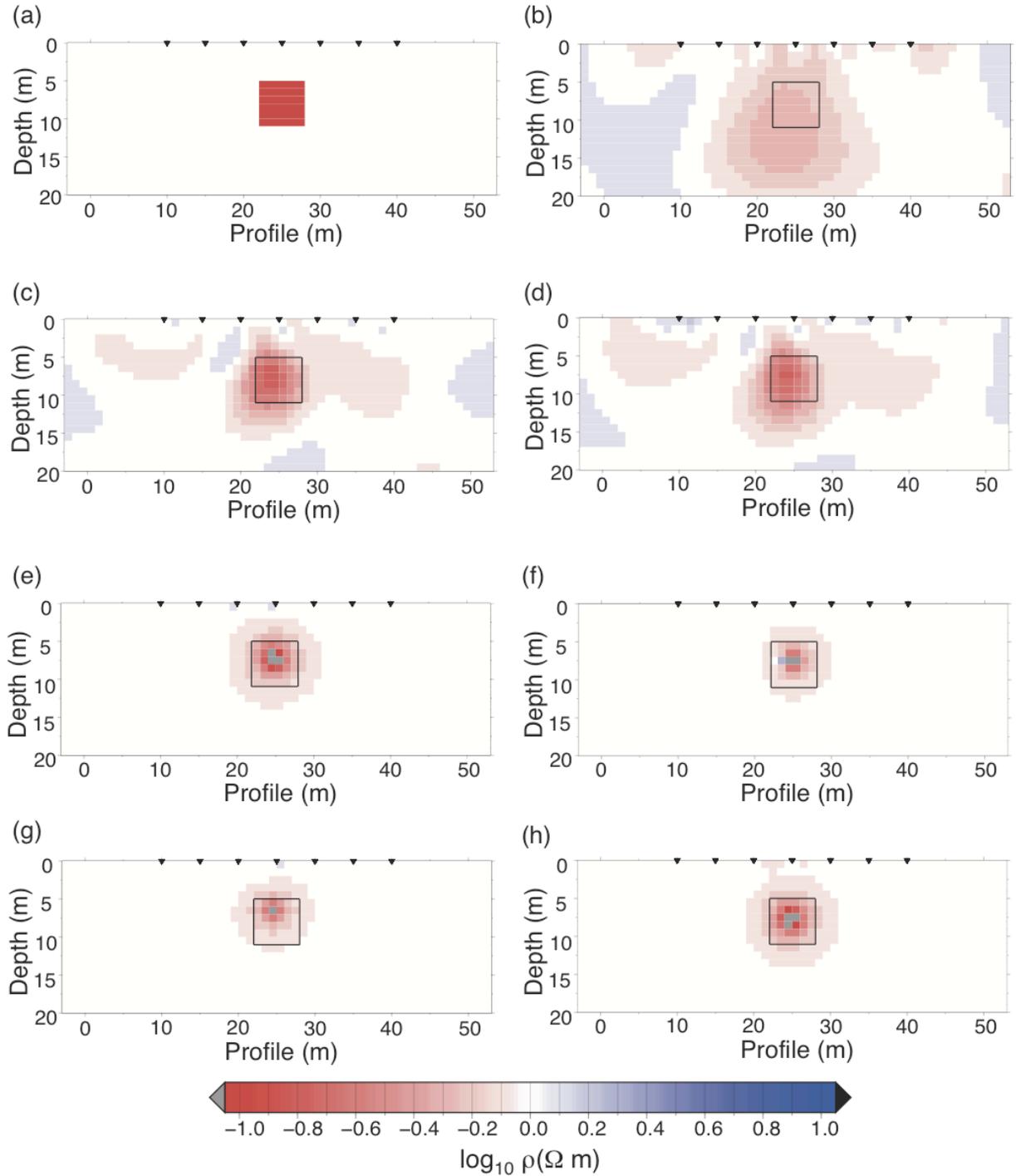

**Fig. 2.** Model differences at $t = 1$ for the shallow prism example. (a) True difference between the synthetic models at $t = 1$ and $t = 0$. Model differences obtained using (b) differencing of Occam inversion models at $t = 1$ and $t = 0$, (c) time-lapse Occam inversion, time-lapse inversion with stochastic regularization using the (d) $l_2$-norm, (e) perturbed $l_1$-norm, (f) minimum support, (g) Cauchy norm, and (h) perturbed $l_1$-norm with negativity constraints applied to the model update. Grey color-coding indicates overestimated differences with respect to the true differences.

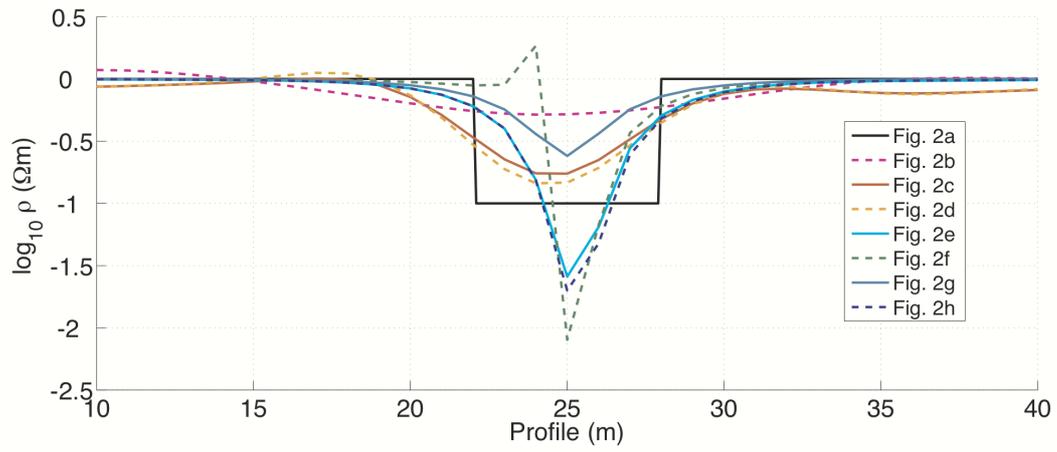

**Fig. 3.** Horizontal slices through the model differences in Fig. 2 at a depth of 8 m.

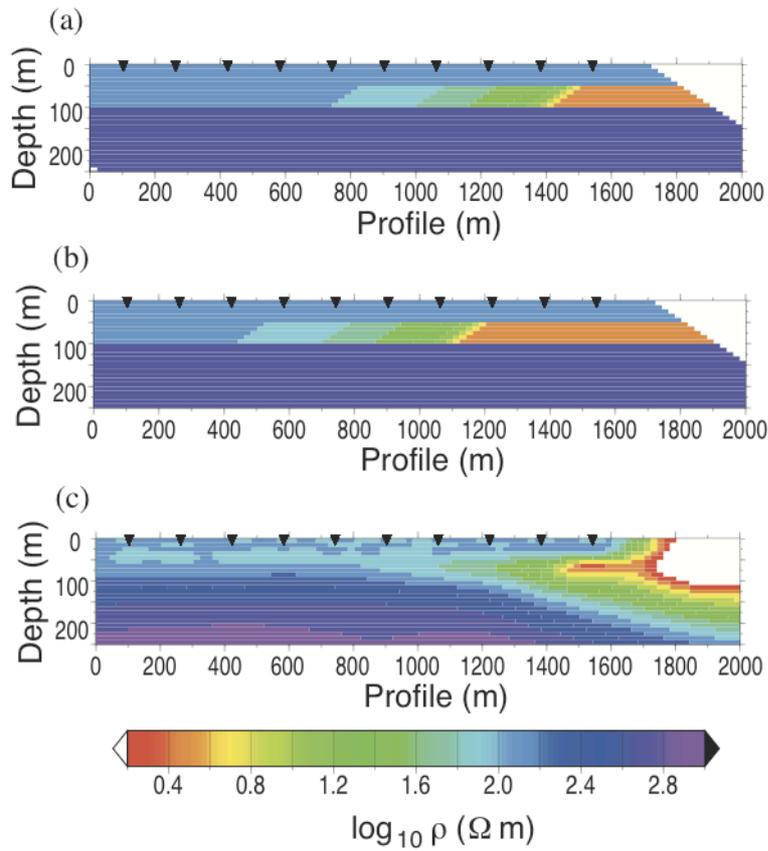

**Fig. 4.** Synthetic 2D models used to generate the data at (a) $t = 0$ and (b) $t = 1$ for the seawater intrusion example. (c) Reference model for the time-lapse inversions obtained by inverting the data at $t = 0$ with an Occam algorithm. The triangles at the top of the figures indicate station locations.

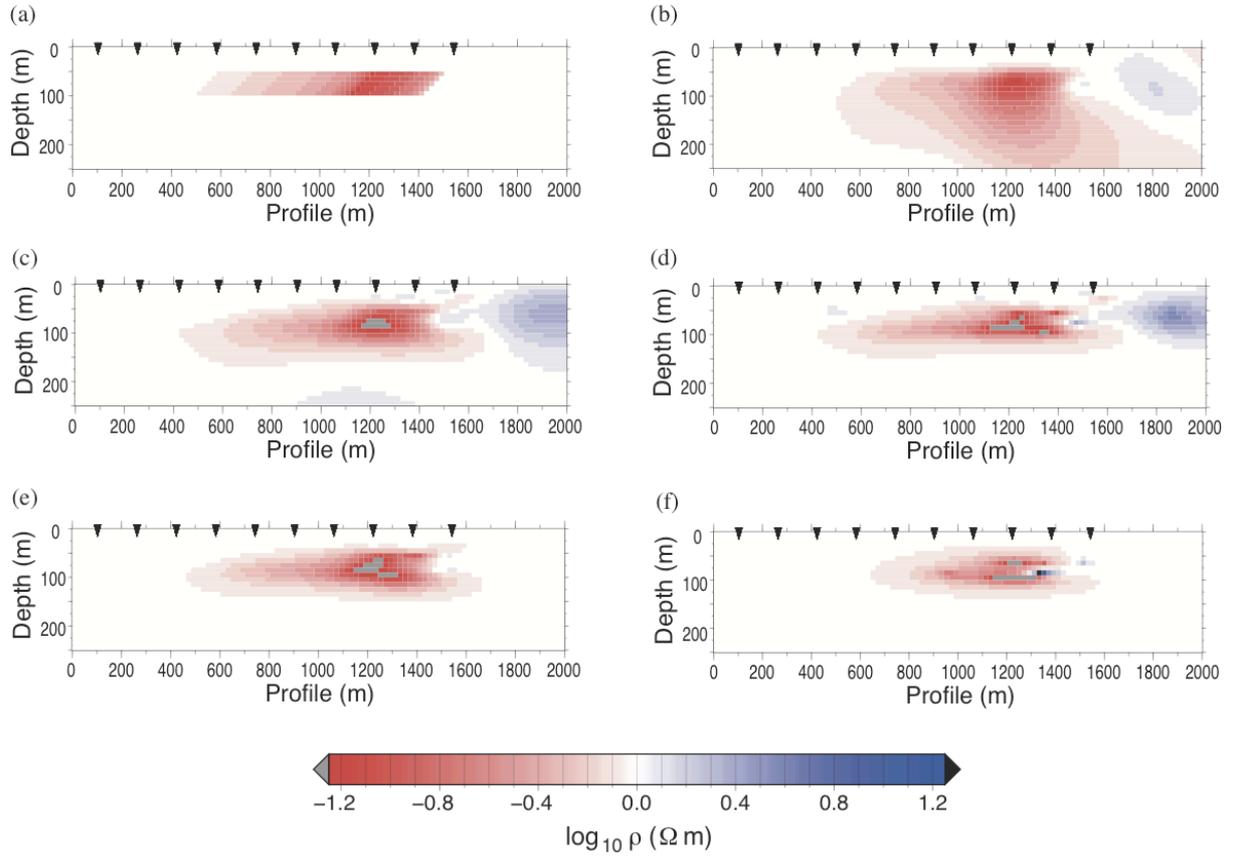

**Fig. 5.** Model differences at $t = 1$ for the seawater intrusion example. (a) True difference between the synthetic models at $t = 1$ and $t = 0$. Model differences obtained using (b) differencing of Occam inversion models at $t = 1$ and $t = 0$, time-lapse inversion with stochastic regularization using the (c) $l_2$-norm, (d) perturbed $l_1$-norm, (e) perturbed $l_1$-norm with negativity constraints, and (f) Cauchy norm with negativity constraints. Grey color-coding indicates overestimated amplitudes with respect to the true differences.

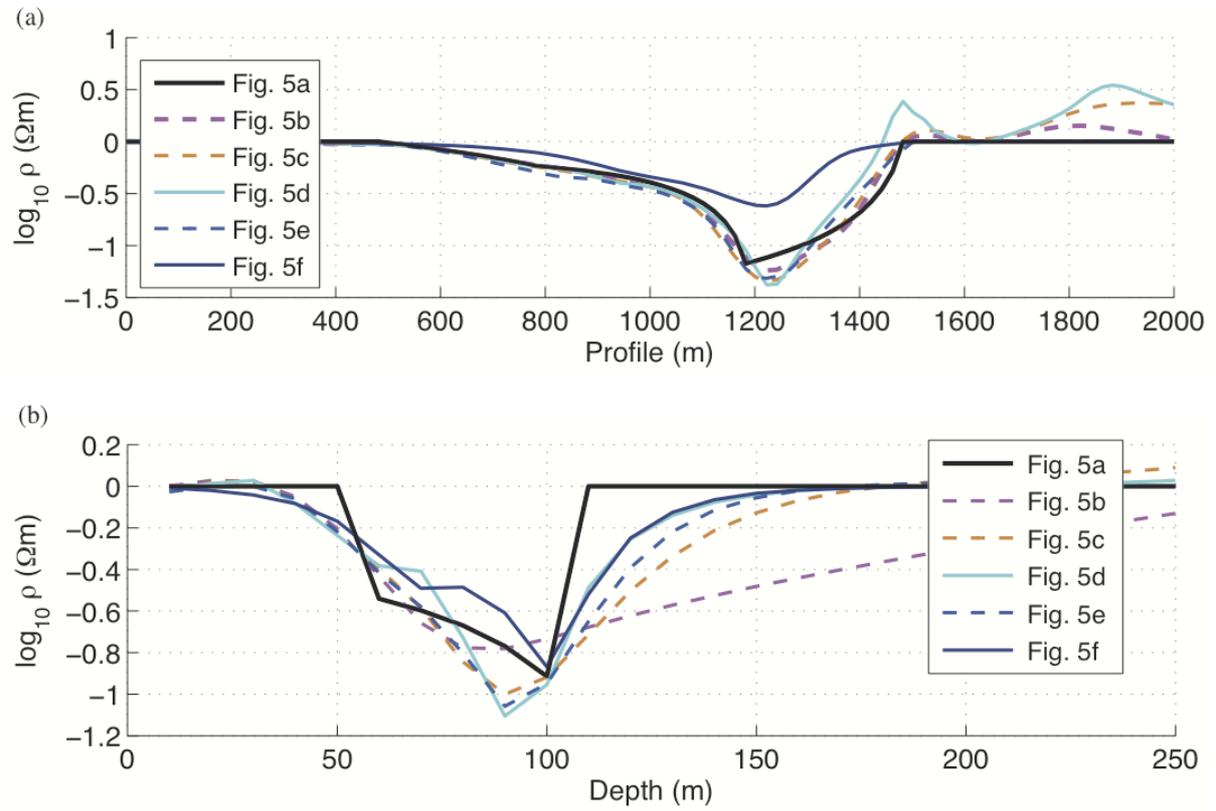

**Fig. 6.** (a) Horizontal and (b) vertical slices of the model differences in Fig. 5 at a depth of 80 m and a profile distance of 1200 m, respectively.